\documentclass[ajl]{emulateapj}

\usepackage{graphicx,epsfig,natbib,color}

\usepackage{graphicx}
\usepackage{epsfig}
\usepackage{natbib}

\begin{document}
\title{Differential Emission Measure Analysis of Multiple Structural Components of Coronal Mass Ejections in the Inner Corona}

\author{X. Cheng\altaffilmark{1,2,3}, J. Zhang\altaffilmark{2}, S. H. Saar\altaffilmark{4}, \& M. D. Ding\altaffilmark{1,3}}

\affil{$^1$ School of Astronomy and Space Science, Nanjing University, Nanjing 210093, China}\email{xincheng@nju.edu.cn}
\affil{$^2$ School of Physics, Astronomy and Computational Sciences, George Mason University, Fairfax, VA 22030, USA}\email{jzhang7@gmu.edu}
\affil{$^3$ Key Laboratory for Modern Astronomy and Astrophysics (Nanjing University), Ministry of Education, Nanjing 210093, China}
\affil{$^4$ Harvard-Smithsonian Center for Astrophysics, 60 Garden Street, Cambridge, MA 02138, USA}

\begin{abstract}

In this paper, we study the temperature and density properties of multiple structural components of coronal mass ejections (CMEs) using differential emission measure (DEM) analysis. The DEM analysis is based on the six-passband EUV observations of solar corona from the Atmospheric Imaging Assembly onboard the \emph{Solar Dynamic Observatory}. The structural components studied include the hot channel in the core region (presumably the magnetic flux rope of the CME), the bright loop-like leading front (LF), and coronal dimming in the wake of the CME. We find that the presumed flux rope has the highest average temperature ($>$8 MK) and density ($\sim$1.0 $\times10^{9}$ cm$^{-3}$), resulting in an enhanced emission measure (EM) over a broad temperature range (3 $\leq$ T(MK) $\leq$ 20). On the other hand, the CME LF has a relatively cool temperature ($\sim$2 MK) and a narrow temperature distribution similar to the pre-eruption coronal temperature (1 $\leq$ T(MK) $\leq$ 3). The density in the LF, however, is increased by 2\% to 32\% compared with that of the pre-eruption corona, depending on the event and location. In coronal dimmings, the temperature is more broadly distributed (1 $\leq$ T(MK) $\leq$ 4), but the density decreases by $\sim$35\% to $\sim$40\%. These observational results show that: (1) CME core regions are significantly heated, presumably through magnetic reconnection, (2) CME LFs are a consequence of compression of ambient plasma caused by the expansion of the CME core region, and (3) the dimmings are largely caused by the plasma rarefaction associated with the eruption.

\end{abstract}
Online-only material: color figures

\section{Introduction}

Coronal mass ejections (CMEs) are perhaps the most spectacular form of solar activity, which expel large quantities of plasma (order of $\sim$10$^{10}$ to $\sim$10$^{13}$ kg) at a speed of hundreds of km s$^{-1}$ with the fastest ones over 3000 km s$^{-1}$ \citep{yashiro04,chen11}. Detailed kinematic analyses find that the acceleration of a CME mainly occurs in the lower corona \citep[e.g., $\leq$ 3.0 $R_\odot$;][]{zhang01}. Subsequently, it propagates into interplanetary space, probably taking the form of a magnetic cloud \citep{burlaga82,klien82}. A magnetic cloud is able to produce severe geomagnetic disturbances if it interacts with the Earth's magnetosphere \citep{gosling93}.

White-light coronagraph observations in the past decades have revealed that many CMEs display a characteristic three-part structure: a bright loop-like leading front (LF), a dark cavity underneath, and an embedded bright compact core \citep{illing83}. When a pre-CME structure lifts off from the associated source region, it can cause the expansion and successive stretching of the overlying magnetic field lines to form a CME. At the same time, the surrounding plasma accumulates at the CME front, thus enhancing the plasma density in the CME LF \citep{cheng11}. In the middle corona (e.g., 3--10 R$_\odot$), densities in the LFs are usually in the order of 10$^{4}$ to 10$^{6}$ cm$^{-3}$, which represent $\sim$10--100 times enhancement over the background corona at those heights \citep{ciaravella03,ciaravella05,schwenn06}. Temperatures of the LFs at 1.5 R$_\odot$ have also been inferred, ranging from 6.0 $\times$ 10$^{3}$ K \citep{ciaravella97} to 2.0 $\times$ 10$^{6}$ K \citep{bemporad07}. The bright cores of CMEs are usually believed to originate from the filament material \citep[the cool but dense plasma suspended in the tenuous corona;][]{gopa06}. Using ultraviolet spectral data, \citet{akmal01} estimated densities in the bright core ranging from 1.4 $\times$ 10$^{6}$ to 7.0 $\times$ 10$^{8}$ cm$^{-3}$ at 1.3 R$_\odot$. Density quickly decreases with increasing height and vary from 1.3 $\times$ 10$^{6}$ to 4.0 $\times$ 10$^{7}$ cm$^{-3}$ at 3.0 R$_\odot$ \citep{raymond04}.

A flux rope structure, involving a set of twisted magnetic field lines around a central axis, is often used to interpret the three-part structure of a CME; for instance, the dark cavity and the bright core of the CME correspond to the whole flux rope and the magnetic dips of the flux rope, respectively \citep[e.g.,][]{low95,chen96,gibson06,riley08}. Such helical flux rope configuration has been reconstructed using nonlinear force-free field models based on photospheric vector magnetogram data \citep[e.g.,][]{canou09,cheng10,guo10,jing10}.

Evidence for the existence of the flux ropes has been found in in-situ solar wind data, which often show a large angle rotation of the magnetic field in magnetic clouds \citep{burlaga82,klien82}. Direct evidence of the flux rope comes in the existence of a conspicuous channel structure in the inner corona before and during a solar eruption \citep{zhang12}. This channel initially appears as a twisted and writhed sigmoidal structure in high temperature passbands, for example in 131 {\AA} at $\sim$10 MK and 94 {\AA} at $\sim$6 MK seen by the Atmospheric Imaging Assembly \citep[AIA;][]{lemen11} on board the \textit{Solar Dynamic Observatory} (\emph{SDO}). The channel evolves toward a semi-circular shape in the slow rise phase and then erupts upward rapidly in the impulsive acceleration phase, producing the front-cavity-core components of the resulting CME \citep{zhang12}. The role that the hot channel plays in the eruption process appears similar to that of flux ropes in the modeling and simulations of CMEs \citep[e.g.,][]{chen96,chen00,lin00,torok05,kliem06,aulanier10,fan07,fan10,olmedo10}.

In addition to the three components, coronal dimmings are another interesting phenomenon with close connections to CMEs. Dimmings can be observed in soft X-rays \citep{sterling97}, in EUV \citep{zarro99,thompson98}, and even in H$\alpha$ passbands \citep{jiang03}. The commonly accepted physical explanation for coronal dimmings is that they represent a density drop in the inner corona resulting from the plasma escape or depletion in the wake of a CME \citep{thompson98,harrison00}, although plasma heating could also play a role, as hot plasma becomes less visible to the instruments sensitive primarily to lower temperatures \citep[e.g.,][]{robbrecht10,cheng11}. Assuming that EUV emission lines are optically thin and temperatures of dimmings do not change significantly, \citet{jin09} estimated a density depletion of $\sim$50\% at the early stage of dimmings. Later on, the intensity of dimming region gradually recovers, with several possible causes identified, including heating of confined plasma in coronal loops \citep{mcintosh07}, interchange reconnections between open magnetic field and small coronal loops \citep{attrill08}, or outflows from the transition region \citep{jin09}.

Previous studies have revealed the properties of CME structures and associated dimmings to a certain extent. Nevertheless, detailed information on the density and temperature properties of these structures is still lacking. Recently, differential emission measure (DEM) analysis has been applied to diagnose the physical properties of a CME. Using \emph{Hinode}/EIS spectroscopic observations, \citet{landi10} reconstructed the DEM distribution of the CME core and found the plasma in the CME core was heated slightly during the CME eruption. From the DEM maps derived by analyzing four \emph{SOHO}/EIT bandpasses, \citet{zhukov04} found that the temperature of the EUV dimmings mainly varied between $\log T\sim$5.0 and $\log T\sim$6.5. Before and after the dimmings, the average DEM level decreased without a change in the overall temperature distribution \citep[also see,][]{tian12}. Moreover, the DEM method has also been used to determine the three-dimensional density and temperature structures in the quiet Sun \citep{vasquez10} and to investigate the temperature evolution in the post-flare loop systems \citep{reeves09}.

In this paper, we apply the DEM method to the latest \emph{SDO}/AIA data, which provides an opportunity for making a significant improvement in understanding CME structures. The detailed thermal properties of multiple CME components, including the flux ropes (seen as hot channels in AIA observations), the bright LFs, and the dimmings are analyzed. Instrument and data reduction are presented in Section 2. In Section 3, we show the results, followed by discussion and conclusions in Section 4.

\section{Instrument and Data Reduction}

\subsection{Instrument}

\begin{figure} 
     \vspace{-0.0\textwidth}    
     \centerline{\hspace*{0.00\textwidth}
               \includegraphics[width=0.7\textwidth,clip=]{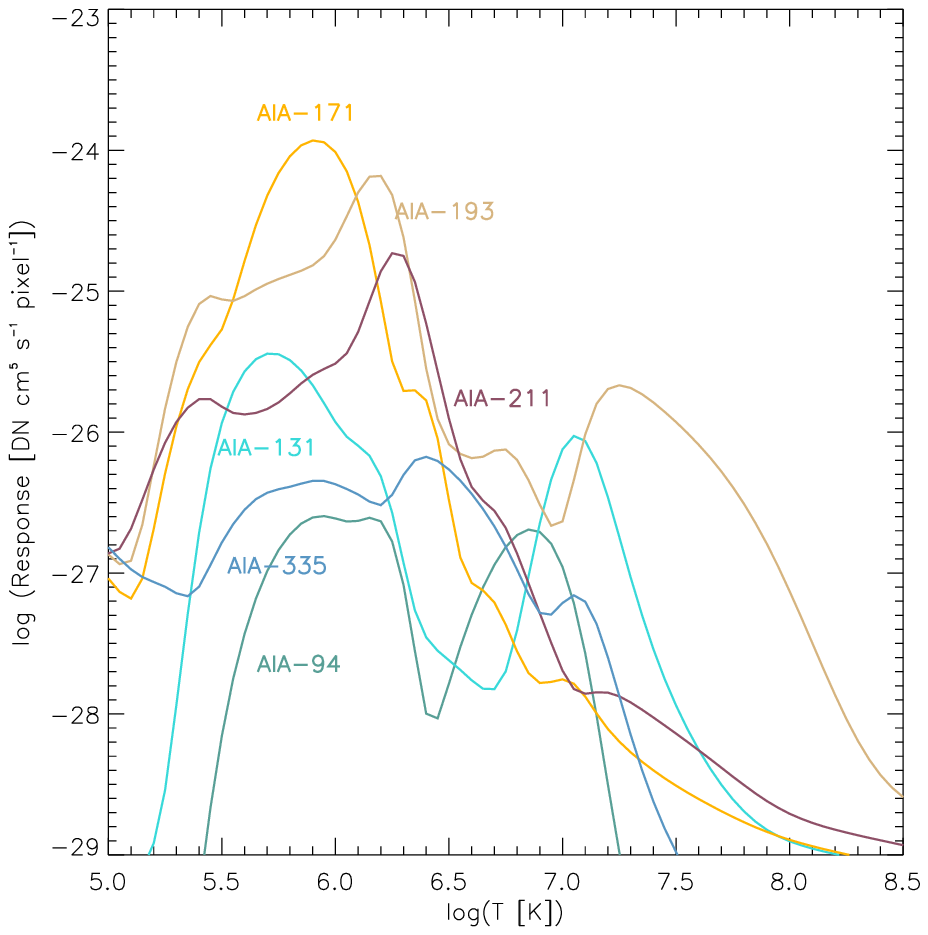}
               }
\vspace{0.0\textwidth}   
\caption{AIA instrument temperature response curves for the six coronal passbands: 131 {\AA}, 94 {\AA}, 335 {\AA}, 211 {\AA}, 193 {\AA}, and 171 {\AA}.} \label{f1}
\end{figure}

The AIA on board \emph{SDO} images the solar atmosphere through ten passbands almost simultaneously, with a temporal cadence of 12 s, a spatial resolution of 1.2\arcsec, and a field of view (FOV) of 1.3$R_\odot$. Six of the filters cover EUV lines formed at coronal temperatures at 131 {\AA} (Fe VIII, Fe XX, Fe XXIII), 94 {\AA} (Fe XVIII), 335 {\AA} (Fe XVI), 211 {\AA} (Fe XIV), 193 {\AA} (FeXII, FeXXIV), and 171 {\AA} (Fe IX), respectively. The temperature response functions of these passbands, as shown in Figure \ref{f1}, indicate an effective temperature coverage from 0.6 to 20 MK \citep{odwyer10,lemen11}. During a solar eruption, the 131 {\AA} and 94 {\AA} passbands are sensitive to the hot plasma from eruption core regions, while the other passbands are better at viewing the cooler LFs and dimming regions \citep[e.g.,][]{cheng11,zhang12}. The multi-passband broad-temperature capability of AIA makes it ideal for constructing DEM models of the distinct CME structures.

\subsection{Method}

The observed flux $F_{i}$ for each passband can be determined by:
\begin{equation}
F_{i}=\int R_{i}(T)\times DEM(T) dT. \\
\end{equation}
where the $R_{i}(T)$ is the temperature response function of passband $i$, and $DEM(T)$ denotes the plasma DEM in the corona. In this work, we use the ``xrt\b{ }dem\b{ }iterative2.pro" routine in SSW package to compute the DEM. This code was originally designed for Hinode/XRT (X-ray Telescope) data \citep{golub04,weber04}, and here is modified slightly to work with AIA data \citep[see also][]{schmelz10,schmelz11a,schmelz11b,wine11}. For more details and tests of this method, see the Appendix.

\subsection{Data Analysis}

\begin{figure} 
     \vspace{-0.0\textwidth}    
     \centerline{\hspace*{0.00\textwidth}
               \includegraphics[width=0.95 \textwidth,clip=]{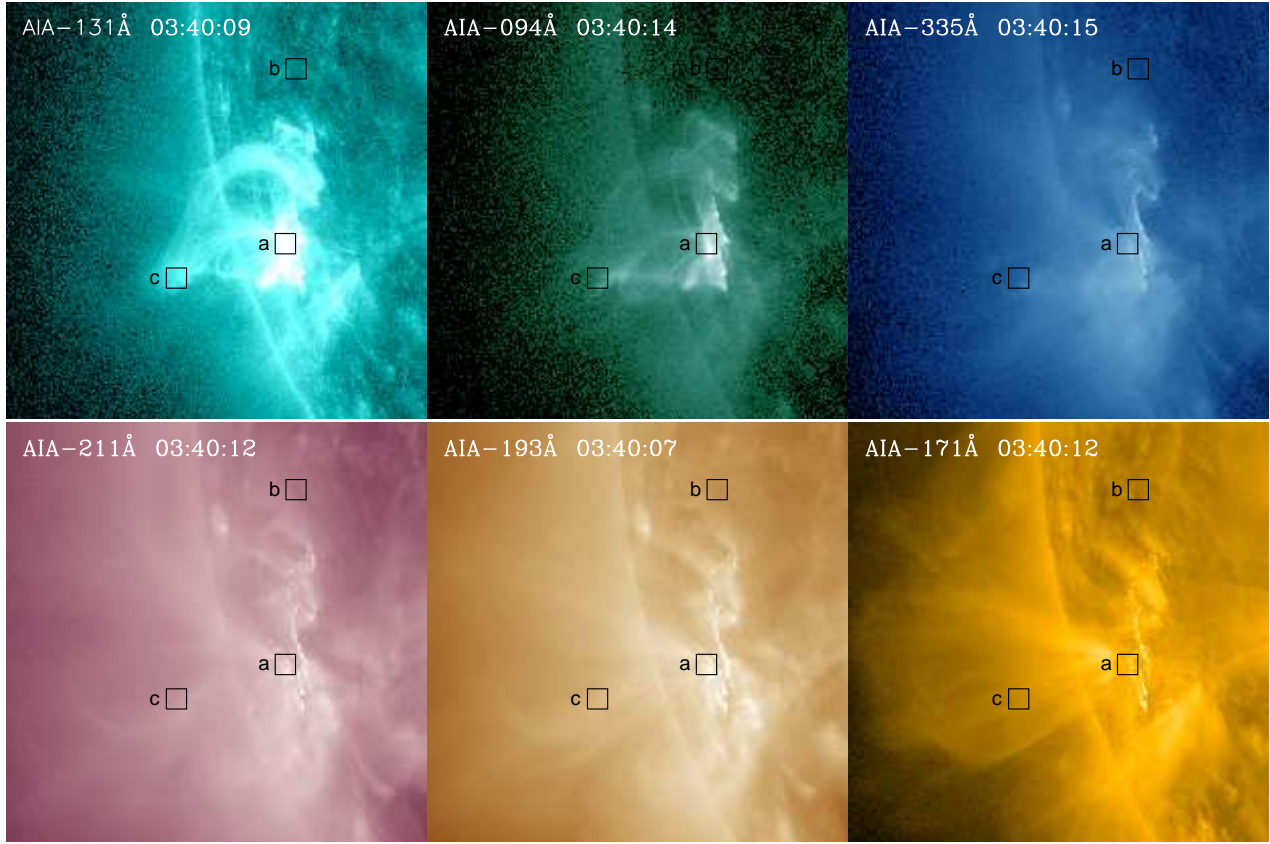}
               }
\vspace{0.0\textwidth}   
\caption{AIA 131 {\AA}, 94 {\AA}, 335 {\AA}, 211 {\AA}, 193 {\AA}, and 171 {\AA} images of the solar eruption on 2011 March 8. The sub-regions used for further analysis, indicated by boxes a, b, and c, are the flare region, quiet-Sun region, and flux rope region, respectively.} \label{f2}
\end{figure}

In this paper, we analyze three well-observed CME events, which occurred on 2010 November 03, 2011 March 08, and 2011 March 07, respectively. We use the CME event on 2011 March 8 as shown in Figure \ref{f2} to illustrate our analysis process. First, we use the ``aia\b{ }prep.pro" routine to process the AIA images in six EUV passbands to 1.5-level, which guarantees a relative coalignment accuracy less than 0.6\arcsec \citep{aschwanden11_solarphy}. Then, we outline three distinct regions (shown as boxes in Figure \ref{f2}) to compute the DEM. Regions a and b correspond to the flare region and the quiet-Sun region, respectively, while a portion of the hot channel is selected in region c. In each region, the DN counts in each of the six passbands are normalized by the exposure time and spatially averaged over all pixels in the region. We use these averaged count rates as the input of ``xrt\b{ }dem\b{ }iterative2" routine to calculate the DEM curve.

\begin{figure} 
     \vspace{-0.0\textwidth}    
     \centerline{\hspace*{0.00\textwidth}
               \includegraphics[width=1.\textwidth,clip=]{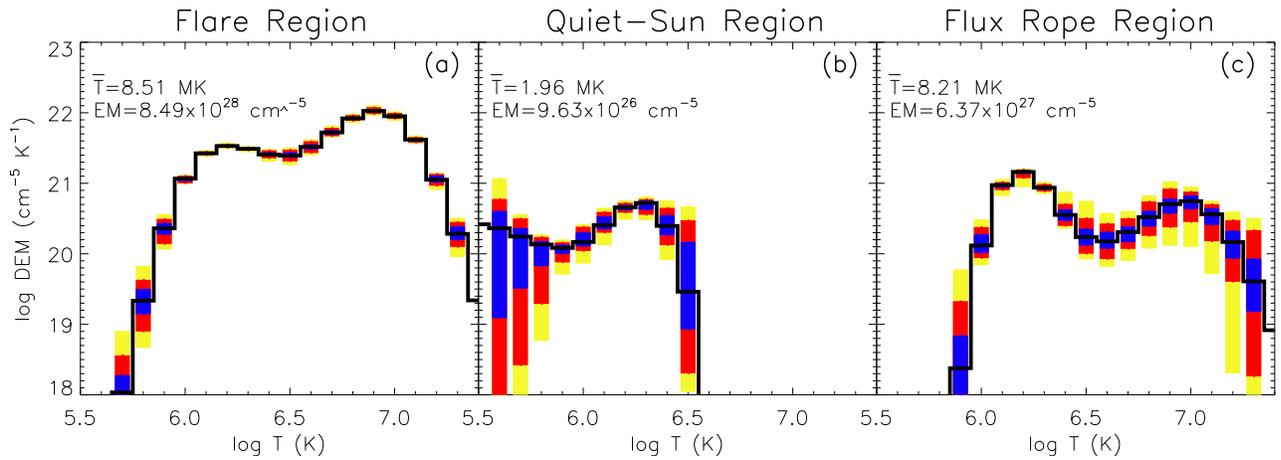}
               }
\vspace{0.0\textwidth}   
\caption{DEM curves for the flare region, quiet-Sun region, and flux rope region of 2011 March 8 CME, whose positions are shown in Figure \ref{f2}. The black solid lines are the best-fit DEM distributions. The blue rectangle represents the region that contains 50\% of the MC solutions. The two red rectangles, above and below the blue rectangle, and the blue rectangle compose the region that covers 80\% of the MC solutions. All of colored rectangles form the region containing 95\% of the MC solutions. Note that, to derive total EM, the DEM is integrated over the temperature range of $\log T$=5.9 to $\log T$=7.3.} \label{f3}
\end{figure}

Figure \ref{f3}(a) shows the DEM result for the flare region. The black solid curve indicates the best-fit DEM solution to the observed fluxes. In order to estimate DEM uncertainties, we compute 100 Monte Carlo (MC) realizations of the data. For each MC simulation, the observed flux $F_{i}$ in each passband is perturbed by an amount of $\delta$, which is randomly drawn from Gaussian distribution with a sigma equal to the uncertainty in the observed flux. The uncertainty is obtained by ``aia\b{ }bp\b{ }estimate\b{ }error.pro" routine (Boerner 2012, private communication). The DEM code is then rerun for each of 100 MC realizations. The 100 MC solutions thus represent 100 ``equivalent" solutions of the original data within the noise on each channel. We use a blue rectangle, as show in Figure \ref{f3}(a), to represent the region surrounding the best-fit solution that contains 50\% of the MC solutions. The region consisting of two red rectangles and a blue rectangle covers 80\% of the MC solutions. The region including all of colored rectangles contains 95\% of the MC solutions. Thus, the upper and lower ends of these colored rectangles can be regarded as estimates of the uncertainties in the best-fit solution, indicating how well the DEM is determined at a given temperature bin.

The flare region DEM shows a broad temperature distribution from $\sim$0.8 MK to 20 MK (or 5.9 $\leq \log T \leq$ 7.3), indicating that plasmas with a wide range of temperatures are present in the compact flare region. Here, we introduce a useful parameter that characterizes the overall temperature of the plasma, i.e., the DEM-weighted average temperature defined as:
\begin{equation}
\bar{T}= \frac{\int DEM(T) \times T dT}{\int DEM(T) dT}. \\
\end{equation}
Using this definition, the average plasma temperature in the flare region is $\sim$9 MK, indicating that high temperature plasma dominates the emission of the flare region. Further, the errors of the DEM solutions are very small in the temperature range of 5.9 $\leq \log T \leq$ 7.3, indicating that the DEM is well constrained by the AIA data over most of the temperature range of the flare region.

The DEM profiles for the selected quiet-Sun region and the flux rope region are shown in Figure \ref{f3}(b) and (c), respectively. We can see that the emission in the quiet-Sun region is dominated by plasma with lower temperatures (5.9 $\leq \log T \leq$ 6.5), over which the DEM is well constrained; the quiet-Sun region has an average temperature at $\sim$2 MK. The peak DEM of the quiet-Sun region is almost one order of magnitude lower than that of the flare region. The flux rope region seems to have two plasma components: the lower temperature component is similar in temperature distribution to that of the quiet region, but there is an additional, well-separated component of high temperature plasma. The average temperature of the flux rope is $\sim$8 MK, indicating that it is somehow significantly heated during the eruption process.

We also calculate the total emission measure (EM) using:
\begin{equation}
EM= \int DEM(T) dT.\\
\end{equation}
It is noteworthy that the total EM of the flare region is $\sim$10$^{29}$ cm$^{-5}$, almost two orders higher than that of the quiet-Sun region of 10$^{27}$ cm$^{-5}$. Due to the contribution of its high temperature plasma, the total EM of the flux rope is up to $\sim$10$^{28}$ cm$^{-5}$, one order higher than that of the quiet-Sun region although lower than that of the flare region. These total EM values, as well as the DEM-weighted average temperature are also indicated in Figure \ref{f3}. Note that all above integrations are carried out over the same temperature range 5.9 $\leq \log T \leq$ 7.3.

\subsection{Uncertainties}

It is well known that DEM inversion is ill-posed and technically fraught with perils. On the one hand, errors in DEM inversion arise from the uncertainties in the response function $R_{i}(T)$, including non-ionization equilibrium effects, non-thermal populations of electrons, modifications of dielectronic recombination rates owing to finite density plasmas \citep[e.g.,][]{summers74,badnell03}, and even radiative transfer effects \citep{judge10}. Moreover, the filling factor of the plasma is unknown, affecting density determinations. Considering these effects, \citet{judge10} estimated an uncertainty of 20\% for $R_{i}(T)$ although it is still a lower limit.

On the other hand, errors in DEM inversion also originate in the uncertainties in the background determination, which is very important for DEM analysis \citep[e.g.,][]{aschwanden11}. In order to obtain the true DEM distribution inside the flux rope, the emission from the background needs to be removed from the observed flux. For flux ropes, the background is determined from the nearby quiet-Sun regions (white boxes in Figure \ref{f4}), which are close to and have the same heliocentric distance (to ensure a similar path along the line of sight in the corona) as the selected flux rope sub-regions. We further inspect the effects of different backgrounds on our results and find that the DEM profiles do not change significantly, but the resulting parameters, e.g., average temperature, total EM, and density, vary by $\sim$30\%. For the CME LFs and dimming regions, however, we use the observed fluxes to calculate the DEM directly without subtracting the flux from a nearby region, since the emission in the LFs and dimmings comes from a large region along the line of sight. In order to reveal the DEM changes, we compare the DEM in the same location before and after the LFs and dimmings formation.

Finally, we note that a cool line component was missing in the old response function of AIA 94 {\AA} passband \citep{foster11,aschwanden11,schmelz11b} and there were also concerns about the accuracy of the 131 {\AA} response function \citep{schmelz11b}. We use the revised response functions (updated on 2012 January 30), in which the cool lines have been added to 94 and 131 {\AA} passbands, to avoid the issue of the missing cool line component.

\section{Results}

\subsection{DEM of CME Flux Ropes}

As previously noted, a CME flux rope usually appears as an isolated channel structure in 131 {\AA} and 94 {\AA} images \citep{cheng11,zhang12}. Three such flux ropes are shown in Figure \ref{f4}. They are viewed in different orientations: the first one is largely seen along its axis, while the other two are mostly viewed from the side. For each event, we select three different sub-regions along the flux rope to calculate their DEMs (black boxes in Figure \ref{f4}); the corresponding background regions are shown by the white boxes in Figure \ref{f4}. The three sub-regions for the 2010 November 03 CME flux rope are shown in the upper left panel of Figure \ref{f4}: one located close to the rim of the flux rope (region a), the other two close to the center of the flux rope (regions b and c). The resulting DEMs are shown in Figure \ref{f5}. We find that at the rim of the flux rope, the plasma has an average temperature of $\bar{T} \sim$8.3 MK, while at the center of the flux rope, the plasma temperature is $\bar{T} \sim$9.0 MK. The average temperature appears to decrease from the center to rim of the flux rope. But, the difference may not be taken seriously given the large uncertainty in these values.

Based on the total EM of the flux rope, we can estimate its density, assuming that the depth of the flux rope along the line of sight is approximately equal to its width, which can be measured directly (as shown by the black lines in Figure \ref{f4}). We calculate the density $n$ in the flux rope using:
\begin{equation}
n= \sqrt{EM/l}.\\
\end{equation}
where $l$ is the depth (or width) of the flux rope. Note that, the filling factor in the density calculation is assumed to be 1. At the rim of the flux rope, with the estimated width of $\sim$40 Mm, the total EM of 2.1 $\times10^{27}$ cm$^{-5}$ corresponds to a density of 7.2 $\times10^{8}$ cm$^{-3}$ (region a). Toward the flux rope center, the density increases to $\sim1.3 \times10^{9}$ cm$^{-3}$, which is mainly due to the enhanced total EM (regions b and c). These results imply that the flux rope has not only a temperature structure but also a density structure. The maxima of both the temperature and the density occur near the flux rope center, with values tending to decrease in the regions away from there. The average temperature, the density, total EM, as well as the width used in the calculation are also indicated in the left-upper corner of Figure \ref{f5}.

Unlike 2010 November 03 event, the flux rope on 2011 March 08 appears as a semi-circular tube with two footpoints fixed in the photosphere. The middle upper panel of Figure \ref{f4} shows the three small sub-regions we selected for DEM analysis. Due to a lower count rate at the top of the flux rope (region a), the DEM curve is poorly constrained in all but a few temperature bins. Nevertheless, the derived parameters for this sub-region are $\bar{T}\sim$10.9 MK, $n\sim$7.4 $\times10^{8}$ cm$^{-3}$ (Figure \ref{f6}(a)). In contrast, the DEM curves in the two legs of the flux curves are well constrained, with the errors smaller (Figure \ref{f6}(b) and (c)). Most of the emission originates from high temperature plasma (6.6 $\leq \log T \leq$ 7.2; $\bar{T} \sim$9.0 MK). The calculated densities are also higher (1.2--2.4 $\times10^{9}$ cm$^{-3}$).

The flux rope on 2011 March 07 is similar to the 2011 March 08 event in terms of orientation, but is associated with a filament. Most of cool filament material is located at the bottom of the flux rope structure, as shown in the right bottom panel of Figure \ref{f4}. From the DEM results of regions a and c (Figure \ref{f7}), we find that a significant amount of the emission is dominated by plasma with a cooler temperature ($\bar{T} <$7.0 MK), likely due to the presence of much cooler and denser filament material (0.6--1.7 $\times10^{9}$ cm$^{-3}$). Region b is far from the filament region, thus is not contaminated by the cooler chromosphere material. It has a higher average temperature ($\bar{T} \sim$9.0 MK) and a lower plasma density (5.3 $\times10^{8}$ cm$^{-3}$), and has more emission from the high temperature component of the plasma. Thus, we conclude that, except when mixed with a much cooler and denser filament component, a flux rope during its eruption is typically of a structure of hot plasma with an average temperature of $\sim$10.0 MK and a density of 1.0 $\times10^{9}$ cm$^{-3}$.

\begin{figure} 
     \vspace{-0.0\textwidth}    
     \centerline{\hspace*{0.00\textwidth}
               \includegraphics[width=0.95 \textwidth,clip=]{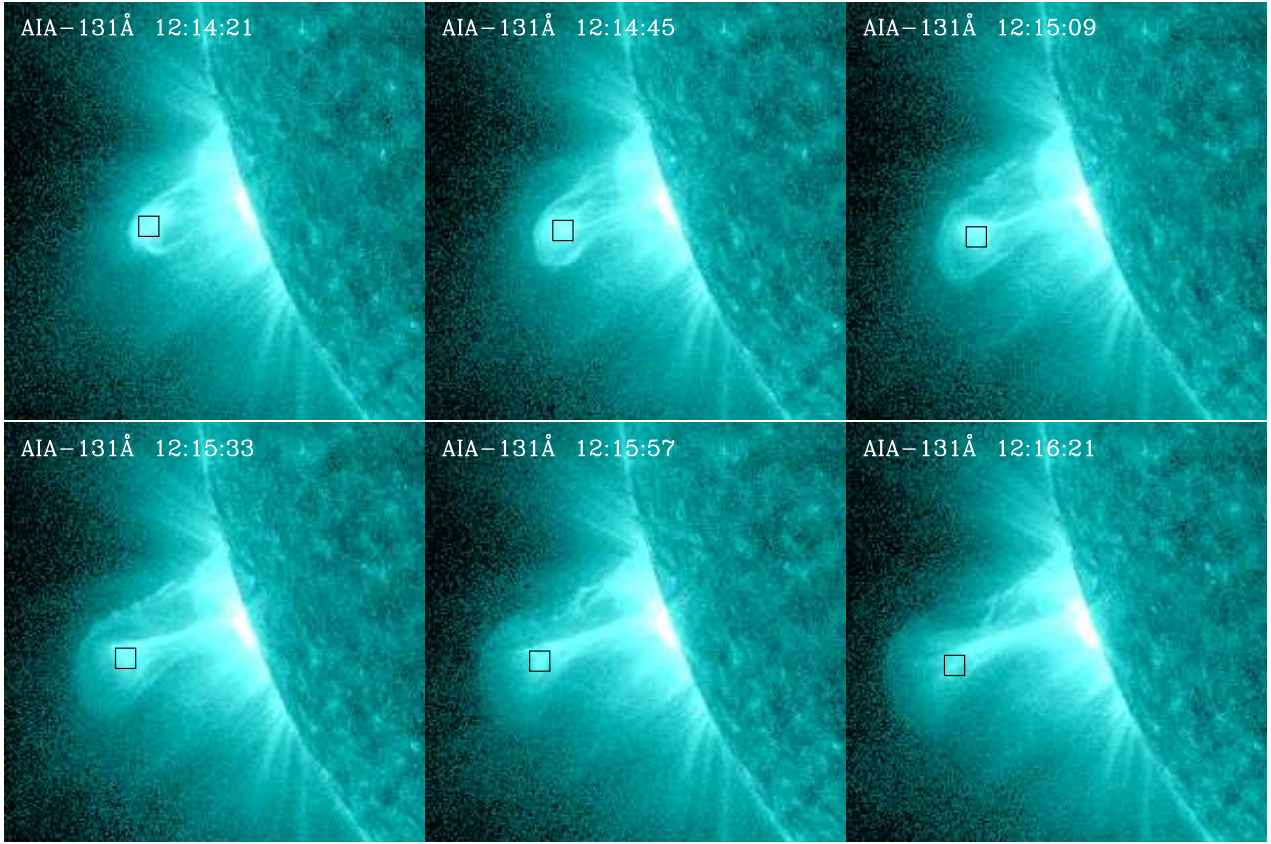}
               }
\vspace{0.0\textwidth}   
\caption{AIA 131 {\AA} images of the CME flux rope that occurred on 2010 November 03. The black boxes denote selected center regions of the flux rope during the eruption.} \label{f8}
\end{figure}

\begin{figure} 
     \vspace{-0.0\textwidth}    
     \centerline{\hspace*{0.00\textwidth}
               \includegraphics[width=0.6 \textwidth,clip=]{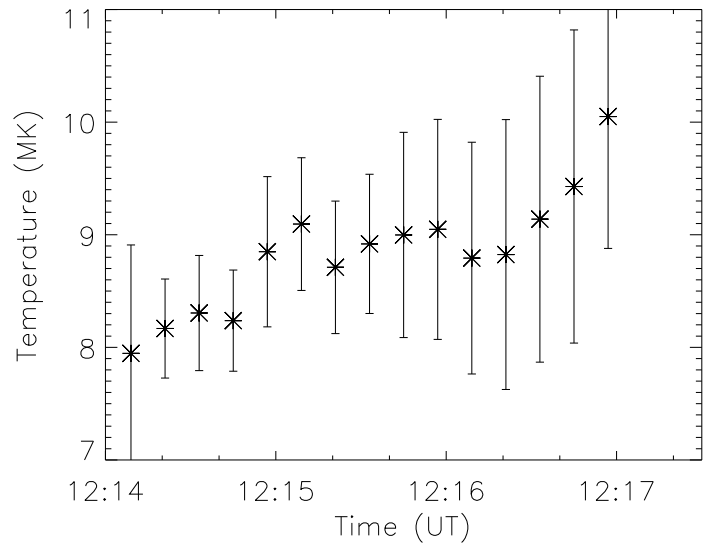}
               }
\vspace{0.0\textwidth}   
\caption{Temporal evolution of the DEM-weighted average temperature in the center of the flux rope for the 2010 November 03 CME as shown in Figure \ref{f8}.} \label{f9}
\end{figure}

The 2010 November 03 flux rope has exceptionally high counts throughout its eruption, thus making it an ideal case for studying the temperature evolution of the flux rope. We select the sub-regions including the maximum intensity as the flux rope centroid, which are indicated by the boxes in Figure \ref{f8}. We derived the DEM distribution every 12 seconds from 12:14 UT to 12:17 UT, and show the evolution of the average temperature in Figure \ref{f9}. We find that the flux rope centroid is further heated during the eruption; i.e., $\bar{T}$ increased from $\sim$8.0 MK to $\sim$10.0 MK, as the flux rope rose up and accelerated; the full kinematic evolution of this event can be found in \citet{cheng11}. Note that, just for the purposes of investigating the relative change of the temperature, the background fluxes were {\it not} subtracted from the observed fluxes before calculating the DEM curves (We implicitly assume the background is small and remains roughly constant.).

\subsection{DEM of CME Leading Fronts}
\begin{figure} 
     \vspace{-0.0\textwidth}    
     \centerline{\hspace*{0.00\textwidth}
               \includegraphics[width=1\textwidth,clip=]{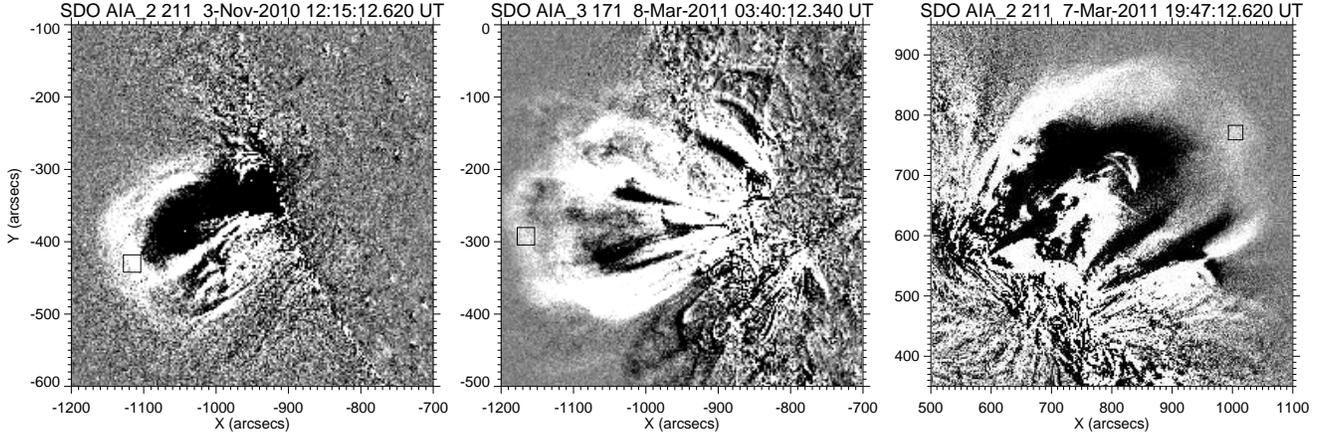}
               }
\vspace{0.0\textwidth}   
\caption{AIA 211 {\AA} (left and right panels) or 171 {\AA} (middle panel) base-difference images of the 2010 November 03, 2011 March 08, and 2011 March 07 CME events. Their base images are taken at 12:00 UT, 03:30 UT, 19:30 UT, respectively. The black box in each panel shows the selected sub-region used to reconstruct the DEM curve of CME LFs.} \label{f10}
\end{figure}

The eruption of the flux rope can push against the overlying magnetic field, whose expansion generates a compression front, observed here as the LF. A LF can be best seen in running or base difference images. Figure \ref{f10} shows the 171 {\AA} or 211 {\AA} base difference images of the three CME events. The base difference images are obtained through subtracting a pre-event image at a fixed time (the base) from the current images. The LF structure in the EUV images is very similar to those in coronagraph images. However, it is not clear whether the brightening front is caused by enhanced plasma density or an increase of temperature (or some combination). We address this issue in this section using DEM analysis.

\begin{figure} 
     \vspace{-0.0\textwidth}    
     \centerline{\hspace*{0.00\textwidth}
               \includegraphics[width=1.0\textwidth,clip=]{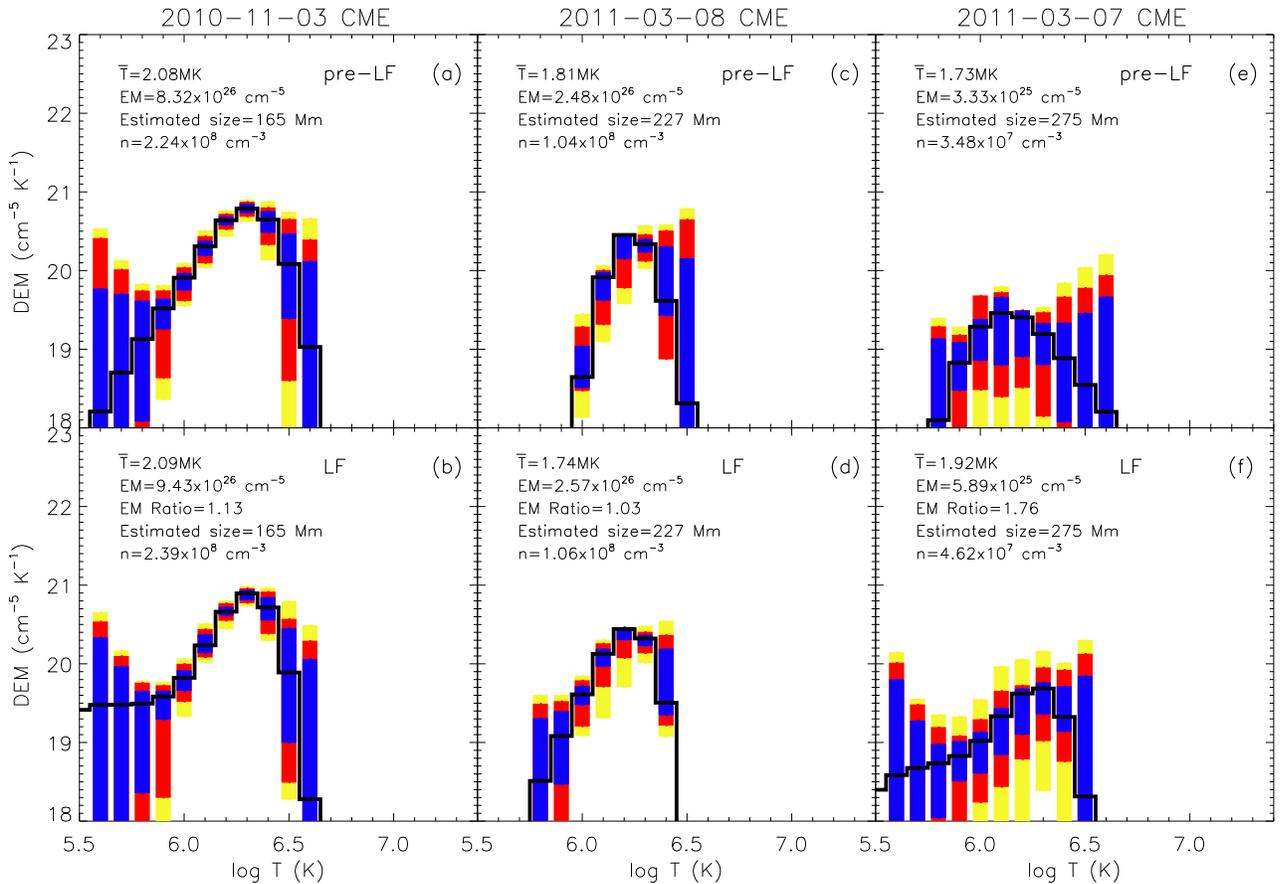}
               }
\vspace{0.0\textwidth}   
\caption{DEM curves of three selected sub-regions in the LFs (shown by the black boxes in Figure \ref{f9}). The upper and bottom panels display the results for the pre-LF and LF, respectively. The black solid lines and the colored rectangles have the same meaning as in Figure \ref{f3}. To derive the total EM, EM ratio, and density $n$, the DEM is integrated over the temperature range of $\log T$=6.0 to $\log T$=6.5.} \label{f11}
\end{figure}

The selected locations for the LFs of the three CMEs are shown in the boxes in Figure \ref{f10}; we calculate the DEM distribution for each region. In order to reveal the DEM changes after the LF formation, the DEM of the same location but for the pre-eruption state (pre-LF) is also calculated. The result of the 2010 November 03 LF is shown in Figure \ref{f11}(b). We find that the most significant (and well constrained) part of the DEM lies only in the lower temperature bins (6.0 $\leq \log T \leq$ 6.4). The DEM indicates that the LF plasma has no significant component with $\log T \geq$6.5.

One interesting result of these analyses is that the shapes of the DEMs do not vary greatly before and after the LF appearance (Figure \ref{f11}(a) and (b)), indicating that the LF has a similar temperature distribution as the pre-eruption state. The average temperature for the LF and pre-LF are both $\sim$2.1 MK. This implies that the LF is not significantly affected by the strong heating process occurring near the core region during the eruption of the flux rope; almost all the released thermal energy, possibly via the process of magnetic reconnection, is confined to the region {\it beneath} the CME LF.

We calculate the EM ratio $R$ to study the variation of the total EM:
\begin{equation}
R= \frac{\int DEM_{\rm LF}(T) dT}{\int DEM_{\rm pre-LF}(T) dT}\\
\end{equation}
where $DEM_{\rm LF}$ and $DEM_{\rm pre-LF}$ represent the DEM of the LF and the pre-LF state (i.e., before the LF forms), respectively. The EM ratio value $R$ is indicated in Figure \ref{f11}(b). It is evident that the EM for the selected region increases up to $\sim$13\% when the LF appears. Since the $\bar{T}$ of LF and pre-LF are similar, this result strongly suggests that the brightening of the LF in EUV passbands is due to enhanced plasma density at the edge of the expanding CME rather than the change of temperature. In other words, the LF is a truly compression front, mainly the result of an enhancement in density. Assuming that the depth of the LF along the line of sight approximates its height from the solar surface (165 Mm), we find that the density increases from $\sim$2.2 to 2.4 $\times10^{8}$ cm$^{-3}$. Note that, the estimated density is an upper limit, since the contribution from the background emission is included in the DEM calculation.

The DEM distribution of the 2011 March 08 CME LF is similar to that of 2010 November 03 CME LF, with most emission coming from low temperatures: 6.0 $\leq \log T \leq$ 6.4 (Figure \ref{f11}(d)). Comparing with the pre-LF region, $\bar{T}$ of the LF does not change but the DEM temperature distribution broadens. Some emission also appears at low temperature ($\log T\sim$6.0) but this is less constrained (Figure \ref{f11}(c) and (d)). These changes result in the EM increasing by $\sim$3\% and the density increasing from $\sim$1.0 to 1.1 $\times10^{8}$ cm$^{-3}$ when the LF appears.

As for the 2011 March 07 CME LF, Figure \ref{f11}(f) shows its DEM result. The DEM is poorly constrained in the range of 6.0 $\leq\log T\leq$ 6.5. For the selected region, the average temperature slightly increases from $\bar{T}\sim$1.7 MK to $\bar{T}\sim$1.9 MK following the LF formation. The EM is enhanced by $\sim$76\% and the density is increased to $\sim$4.6 $\times10^{7}$ cm$^{-3}$, compared to the pre-LF of $\sim$3.5 $\times10^{7}$ cm$^{-3}$ (Figure \ref{f11}(e) and (f)). It is worth mentioning that these values carry a considerable uncertainty, given the large errors in the DEM solutions.

Note that, all the density changes that we derive above include an additional uncertainty from the simple estimation of the LF's depth. Moreover, the density change in the LF depends on sampled different regions, e.g., near the nose or at the flank. Nevertheless, we find that the percentage of the density increase is always less than 50\% in our study, thus providing an upper limit.

\subsection{DEM of CME Dimming Regions}
\begin{figure} 
     \vspace{-0.0\textwidth}    
     \centerline{\hspace*{0.00\textwidth}
               \includegraphics[width=1.\textwidth,clip=]{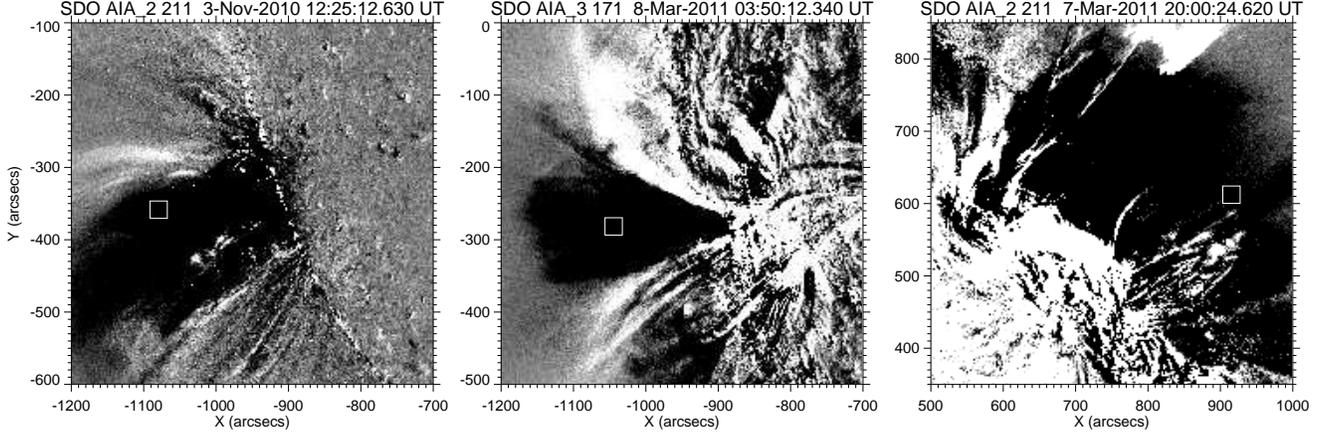}
               }
\vspace{0.0\textwidth}   
\caption{AIA 211 {\AA} (left and right panels) or 171 {\AA} (middle panels) base-difference images showing the dimmings associated with the three CMEs. Their base images are taken at 12:00 UT, 03:30 UT, 19:30 UT, respectively. The white box in each panel indicates the selected dimming region.} \label{f12}
\end{figure}

\begin{figure} 
     \vspace{-0.0\textwidth}    
     \centerline{\hspace*{0.00\textwidth}
               \includegraphics[width=1.0\textwidth,clip=]{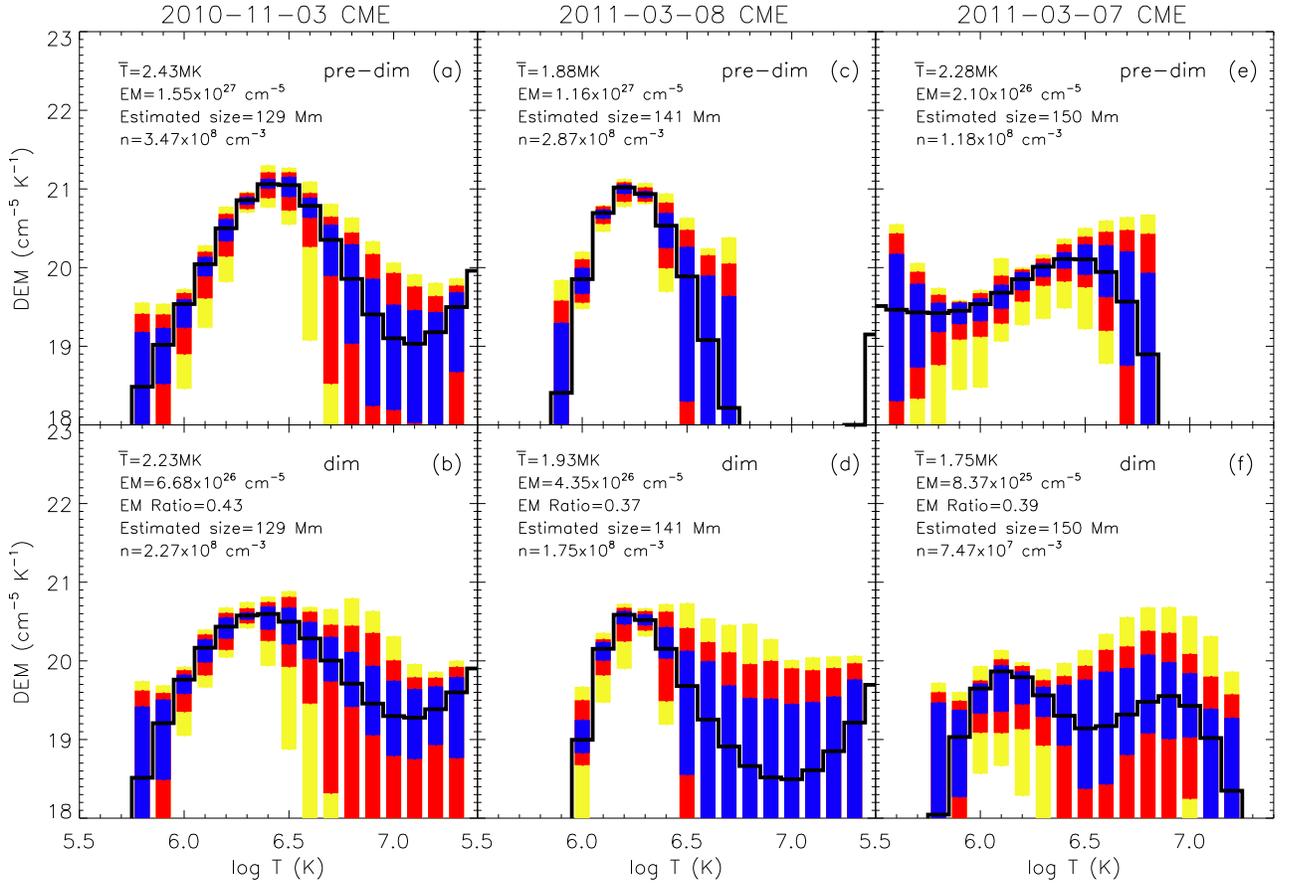}
               }
\vspace{0.0\textwidth}   
\caption{DEM curves of the three selected dimming sub-regions (shown by the white boxes in Figure \ref{f9}). The upper and bottom panels display the results for the pre-dimming and dimming regions, respectively. The black solid lines and the colored rectangles have the same meaning as in Figure \ref{f3}. In order to derive the total EM, EM ratio, and density $n$, the DEM is integrated over the temperature range of $\log T$=6.0 to $\log T$=6.5.} \label{f13}
\end{figure}

The dimming regions caused by 2010 November 03, 2011 March 08, and 2011 March 07 CMEs are shown in Figure \ref{f12}. To investigate the DEM of the dimming regions, we avoid areas that include hot plasma, e.g., the hot flux rope areas in the images. The selected dimming sub-regions are indicated by the boxes in Figure \ref{f12}. We use the same method, comparing their DEM distributions with the same region just before the dimming region appears (pre-D). The results for the three events are shown in Figure \ref{f13}.

For the dimming on 2010 November 03 (Figure \ref{f13}(b)), the well-determined portion of the DEM profile lies primarily in the range of 6.0 $\leq\log T\leq$ 6.6, and $\bar{T} \sim$2.2 MK. The same region prior to the dimming has a very similar temperature, but the DEM peak following the dimming decreases by almost one order of magnitude. The total EM decreases by 57\% (Figure \ref{f13}(a) and (b)). Also assuming that the depth of the dimming along the line of sight is comparable to its height, we estimate that the density decreases from $\sim$3.5 to 2.3 $\times10^{8}$ cm$^{-3}$. Similarly, the DEM of 2011 March 08 dimming is centered around $\bar{T} \sim$1.9 MK with temperatures spanning 6.1 $\leq\log T\leq$ 6.4. The total EM decreases by $\sim$63\% and the density decreases to $\sim$1.8 $\times10^{8}$ cm$^{-3}$ (Figure \ref{f13}(c) and (d)). These results also hold for the 2011 March 07 dimming (Figure \ref{f13}(e) and (f)), with the main DEM distribution spanning from $\log T\sim$6.0 to $\log T\sim$6.8, with $\bar{T}$ at $\sim$1.8 MK, with the decrease of EM by 61\% and of the density to $\sim$7.5 $\times10^{7}$ cm$^{-3}$. Note that the DEM reconstruction here is rather poorly constrained; the errors vary by several orders of magnitude in most of temperature bins. Low counts are the cause of the large uncertainties in the inferred DEM. Nevertheless, we can conclude with high confidence that the dimming is mainly caused by the decrease of plasma density in the region.

\section{Discussion and Conclusions}

We summarize in Table \ref{tb} various properties of different CME structures, including average temperature, width of the DEM curve at the 10\% of peak value (using a single Gauss fitting to the DEM curve), maximum DEM, total EM, and the density. The quantitative results from the DEM analysis further support our previous result based on a qualitative argument: a CME consists of a high temperature flux rope and a cooler LF \citep{cheng11}. By tracking the centroid of the flux rope in the 2010 November 03 event, we find that the DEM-weighted temperature of the flux rope increases as it accelerates outward. Observations of the curve-in of the flux rope legs, the shrinkage of the post-flare loops \citep{cheng11}, and the presence of high energy hard X-ray sources surrounding the flux rope \citep{glesener11,guo12}, taken together, argue that magnetic reconnection taking place in the current sheet underneath the flux rope is responsible for heating the plasma inside the flux rope to high temperature (making it visible at 131 {\AA} and 94 {\AA}). A similar flux rope heating scenario was suggested by \citet{landi10}, who also attributed the heating to magnetic reconnection.

We estimate the density of CME flux ropes during the eruption to be as high as $\sim$1$\times10^{9}$ cm$^{-3}$, similar to that of coronal loops in the lower corona \citep[e.g.,][]{ugarte05,aschwanden08}. This implies that the flux rope must originate in the core field structure of the associated active region, and that the high density of the entire structure is kept throughout the early evolution of the flux rope.

The AIA data are also effective for reconstructing the DEM of CME LFs. Most of the LF plasma is confined to a low-moderate temperature range (6.1 $\leq\log T\leq$ 6.5), with little emission at lower or higher temperatures and a DEM-weighted temperature in the range of $\bar{T} \sim$1.7 MK to 2.1 MK. Since the plasma is prohibited from moving across magnetic field lines, the energy released by magnetic reconnection is difficult to transfer into the CME LF. Therefore, during the magnetic reconnection, the CME flux rope is heated but the LF remains about the same temperature as the quiet coronal loops \citep[e.g., T $\sim$1.0--3.0 MK;][]{schmelz10,schmelz11a,schmelz11b}.

For the LF regions, the enhancement of the overall EM (4\%--76\%) at typical corona temperature indicates that the brightening of the LFs is mainly due to an increase in plasma density. Assuming no depth change along the line of sight when the LF passes, we estimate a density increase of 2\% to 32\% for different LF regions at $\sim$1.2--1.4$R_\odot$. Using polarized brightness Mauna Loa data, \citet{bemporad07} estimated that the density of CME LFs increases about 35\% over the background coronal density at $\sim$1.6$R_\odot$. Using AIA data and the DEM analysis, but assuming no temperature change before and after the LF appearance, \citet{kozarev11} found that the densities increase by $\sim$12\% and $\sim$18\% at $\sim$1.3$R_\odot$ in two different bright fronts, respectively. Generally, our results are consistent with previous estimations for the density change of the CME LFs in the low corona.

Dimming regions show changes in the opposite sense as the LFs. The well-constrained portion of dimming region DEM is in the range of 6.0 $\leq\log T\leq$ 6.5. DEM-weighted temperature varies from 1.7 MK to 2.2 MK, similar to that of the LFs. The DEM-weighted temperature doesn't change much before and after the dimming. However, the peak DEM decreases by as much as a factor of 10, which results in a total EM decrease of $\sim$60\%. Similarly, assuming no depth change along the line of sight before and after the dimming, the decreased EM corresponds to the depletion of the density of $\sim$40\%. This shows that the dimming is mostly caused by the density rarefaction or depletion in the lower corona \citep[also see,][]{thompson98,harrison00,harrison03,zhukov04,jin09,tian12}.

We summarize our main conclusions from the DEM analysis for the three distinct CME structures below.

\begin{figure} 
     \vspace{-0.0\textwidth}    
     \centerline{\hspace*{0.00\textwidth}
               \includegraphics[width=1.\textwidth,clip=]{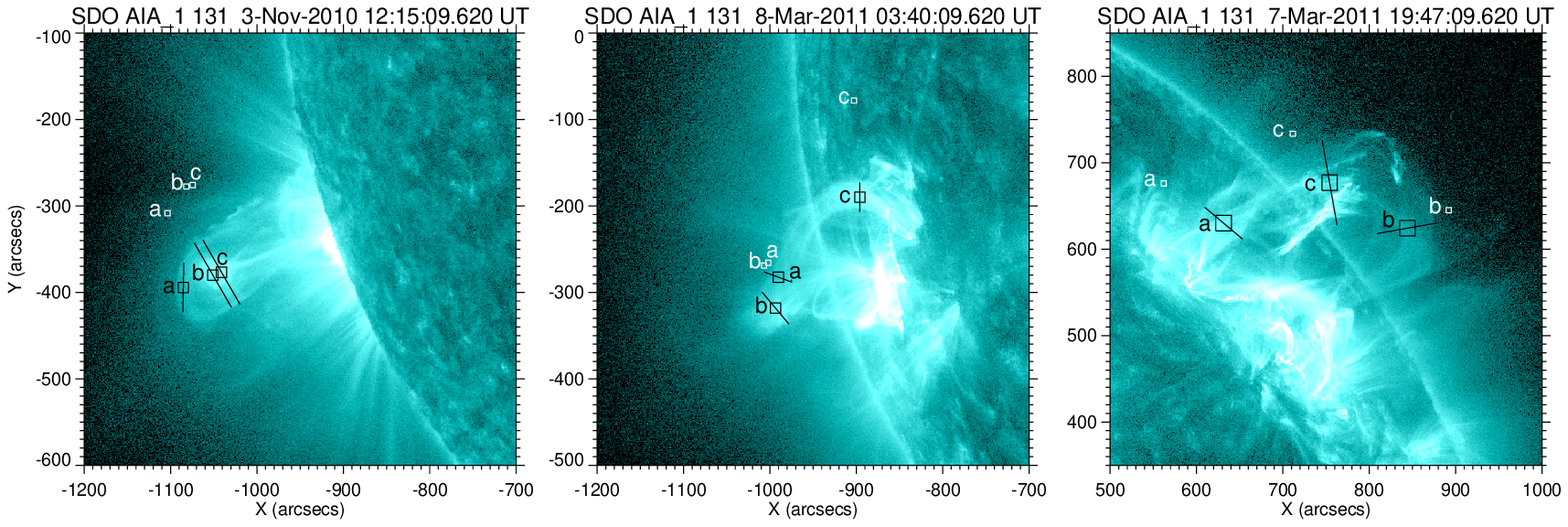}
               }
     \centerline{\hspace*{0.00\textwidth}
               \includegraphics[width=1.\textwidth,clip=]{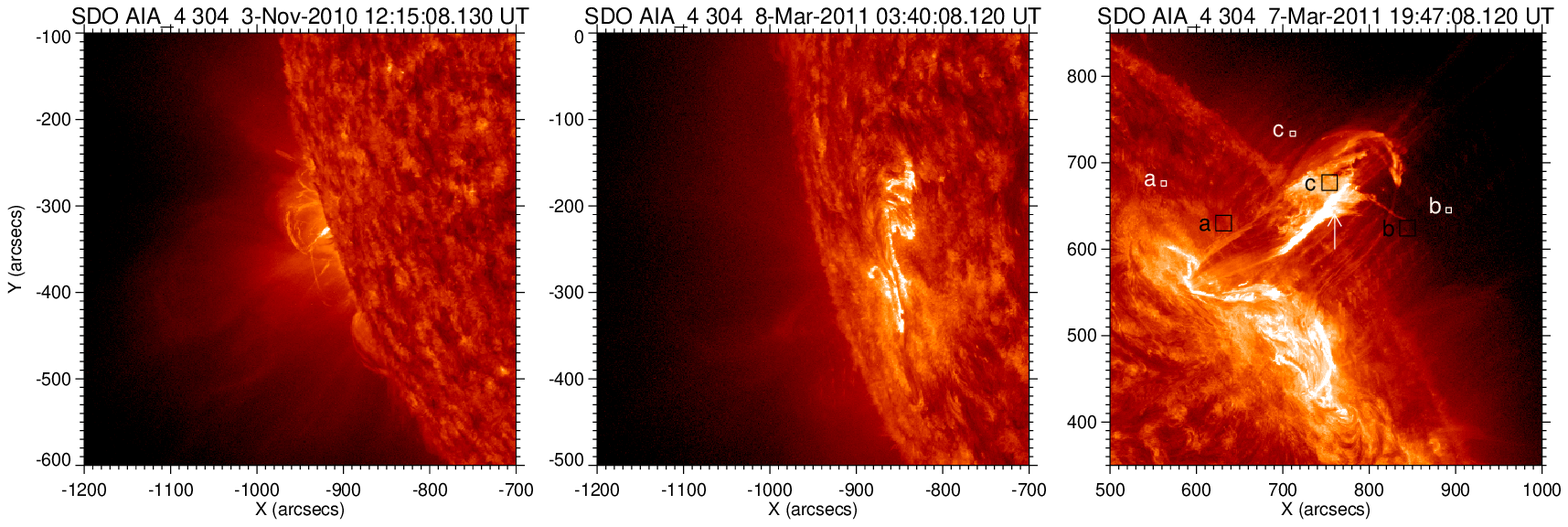}
               }
\vspace{0.0\textwidth}   
\caption{Top: AIA 131 {\AA} images of three CME flux ropes occurring on 2010 November 03, 2011 March 08, and 2011 March 07, respectively. The black boxes (a, b, and c) in each panel show the selected sub-regions used to reconstruct the DEM curves. The sub-regions shown by the white boxes indicate the locations where the background emissions are taken. Bottom: AIA 304 {\AA} images showing the CME flux rope-associated filament on 2011 March 07.} \label{f4}

\end{figure}

\begin{figure} 
     \vspace{-0.0\textwidth}    
     \centerline{\hspace*{0.00\textwidth}
               \includegraphics[width=1.\textwidth,clip=]{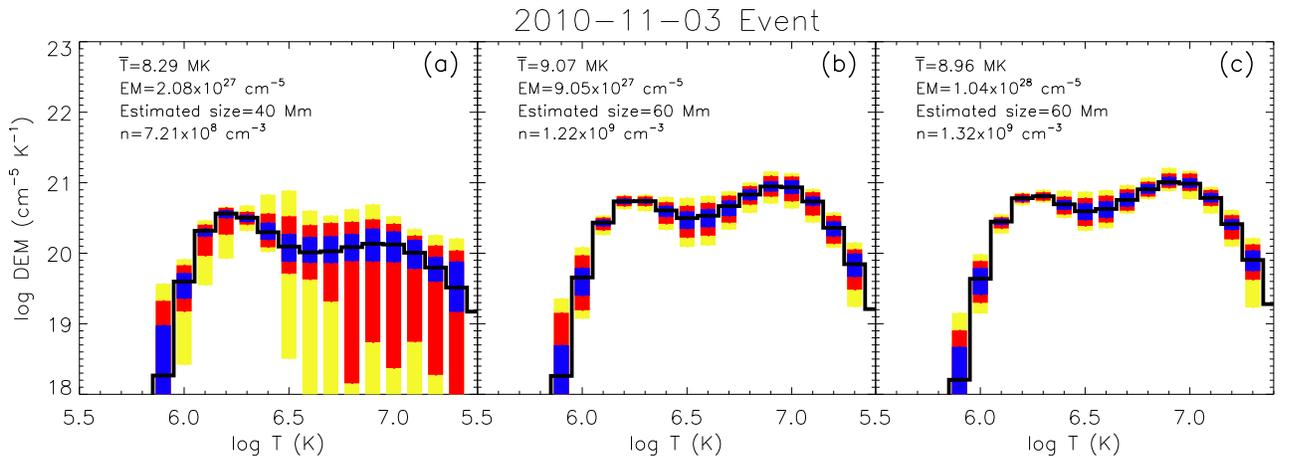}
               }
\vspace{0.0\textwidth}   
\caption{DEM curves of the sub-region a, b, and c of 2010 November 03 flux rope (left upper panel of Figure \ref{f4}). The black solid lines and the colored rectangles have the same meaning as in Figure \ref{f3}. In order to derive the total EM and density $n$, the DEM is integrated over the temperature range of $\log T$=6.0 to $\log T$=7.2.} \label{f5}
\end{figure}

\begin{figure} 
     \vspace{-0.0\textwidth}    
     \centerline{\hspace*{0.00\textwidth}
               \includegraphics[width=1.\textwidth,clip=]{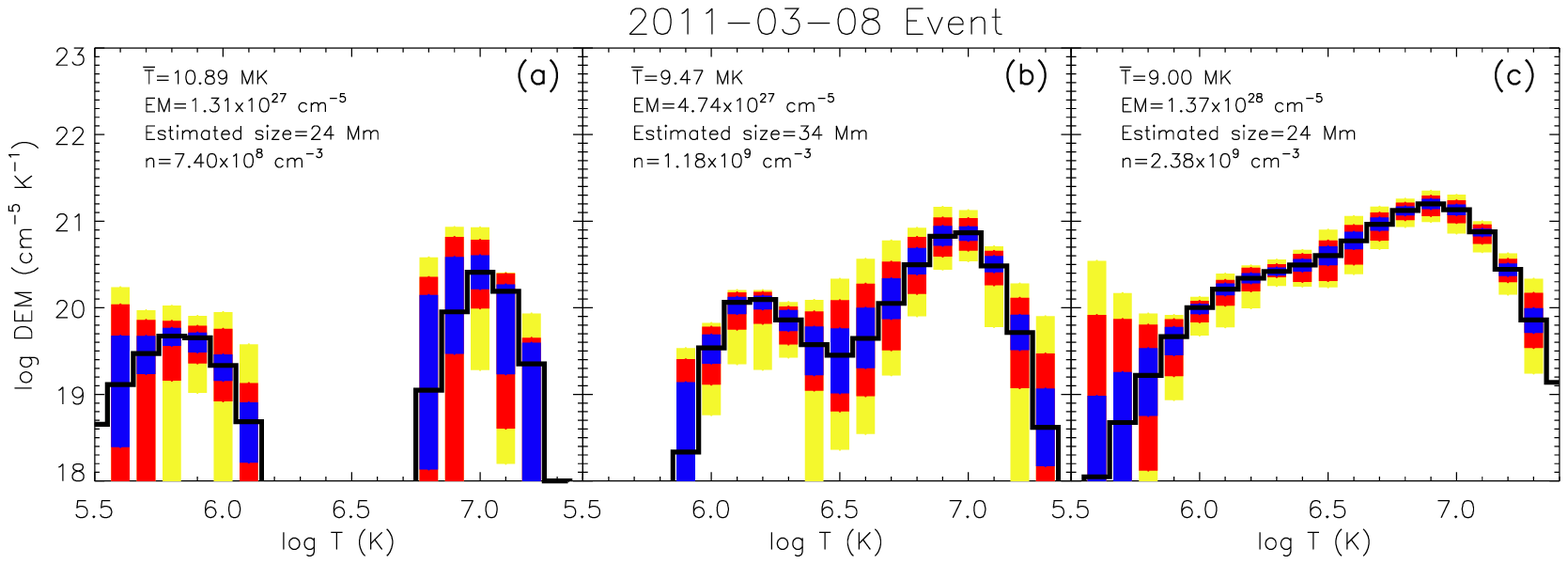}
               }
\vspace{0.0\textwidth}   
\caption{Same as in Figure \ref{f5} but for 2011 March 08 flux rope.} \label{f6}
\end{figure}

\begin{figure} 
     \vspace{-0.0\textwidth}    
     \centerline{\hspace*{0.00\textwidth}
               \includegraphics[width=1.\textwidth,clip=]{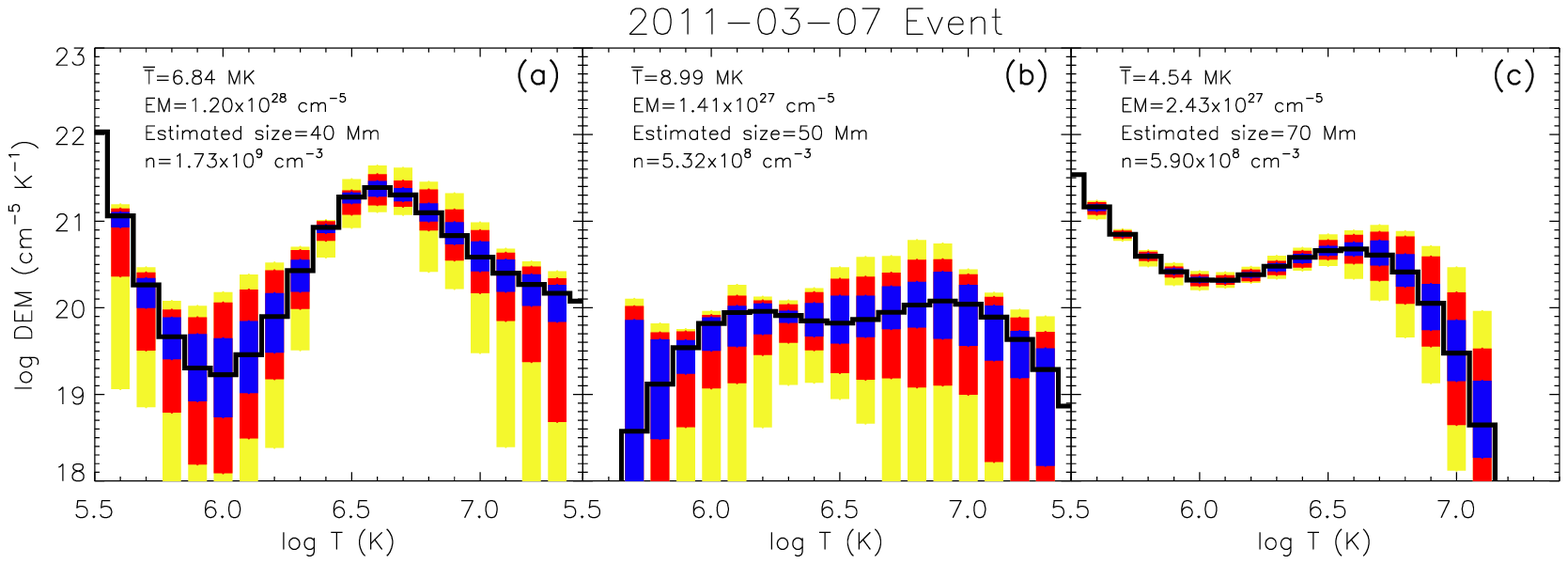}
               }
\vspace{0.0\textwidth}   
\caption{Same as in Figure \ref{f5} but for 2011 March 07 flux rope.} \label{f7}
\end{figure}

1. The plasma in flux ropes generates significant emissions over a broad temperature range of 6.5 $\leq\log T \leq$ 7.3. For three flux ropes studied here, the densities vary from 0.5 $\times10^{9}$ to 2.4 $\times10^{9}$ cm$^{-3}$ and the DEM-weighted average temperatures are all above 8 MK. In one case, the DEM-weighted temperature even increases to $\sim$10 MK as the flux rope rises up, probably due to the continuous magnetic reconnection. The presence of filament material within the magnetic dips of the flux rope can cause cooler {\it apparent} temperatures due to the mixture of hot and cold plasmas along the line-of-sight (e.g., 2011 March 07 event as shown in Figure \ref{f4}).

2. In three selected CME LF regions, the emission mostly comes from the cool plasma (6.1 $\leq\log T\leq$ 6.5), with the DEM-weighted average temperature unchanged by passage of the LF. Comparing to pre-LF regions, the density increases by 2\% to 32\%. We can thus conclude that the brightening of the LFs largely due to the plasma compression at the CME front rather than the increase of the temperature.

3. Cool plasma (6.0 $\leq\log T\leq$ 6.5) also dominates the emission in the dimming regions, again with no change in the DEM-weighted average temperature. The decreased EM, and resulting dimmings, is thus due to density depletion in the low corona. The density reduction for three selected dimming regions is by $\sim$35\% to $\sim$40\%.

In short, DEM analysis appears to be an important tool to diagnose the temperature and density properties for various structural components during CME eruptions. These basic parameters can provide valuable information to guide future CME modeling and simulations.

\acknowledgements
The authors are grateful to the anonymous referee for his/her comments and persistence, which improved the manuscript significantly.
The authors also thank Mark A. Weber for many valuable comments.
SDO is a mission of NASA's Living With a Star Program.
X.C. and M.D.D. are supported by NSFC under grants 10878002 and 10933003 and NKBRSF under grant 2011CB811402.
X.C. is also supported by the scholarship granted by the China Scholarship Council (CSC) under file No. 2010619071.
J.Z. is supported by NSF grant ATM-0748003, AGS-1156120 and NASA grant NNG05GG19G.
S.S. is supported by contract SP02H1701R from Lockheed-Martin to the Smithsonian Astrophysical Observatory for AIA analysis.


\appendix

The ``xrt\b{ }dem\b{ }iterative2.pro" in the SSW package is a DEM reconstruction routine, originally developed by Mark A. Weber \citep[e.g.,][]{golub04,weber04}. It uses a forward-fitting method, in which a DEM profile is guessed and then folded through the response of each passband to produce predicted fluxes. This process is iterated using Levenberg-Marquardt least-squares minimization, until the predicted fluxes are close to the observed ones. The DEM profile is interpolated using N-1 spline functions, representing the degrees of freedom for N different passband observations, which are directly manipulated by the well-known and much-tested ``mpfit.pro" routine from Craig B. Markwardt\footnote{http://cow.physics.wisc.edu/~craigm/idl/idl.html}.

To assess the ability of ``xrt\b{ }dem\b{ }iterative2.pro" to reproduce DEMs, we have simulated AIA observations with several input DEM models, and then applied the forward-fitting routine to simulated AIA data to derive the best-fit DEM solutions. We then compared the reconstructed DEMs with input model DEMs, and explored uncertainties by computing 100 different MC realizations (adding random noise within an uncertainty obtained by ``aia\b{ }bp\b{ }estimate\b{ }error.pro" to the simulated observations) and fitting these as well. The fluctuations of 100 MC solutions measure the confidence in the reconstructed DEMs; typically, lower scatter in 100 MC solutions reflects smaller uncertainty in the reconstructed DEM solution.

The model DEMs that we have tested include the single and double isothermal DEMs, single and double Gaussian DEMs, and an active region DEM from CHIANTI \citep[version 5.2.1;][]{dere97,landi06}. We also discuss some specific representative cases in more detail.

\begin{figure} 
     \vspace{-0.0\textwidth}    
     \centerline{\hspace*{0.00\textwidth}
               \includegraphics[width=0.7\textwidth,clip=]{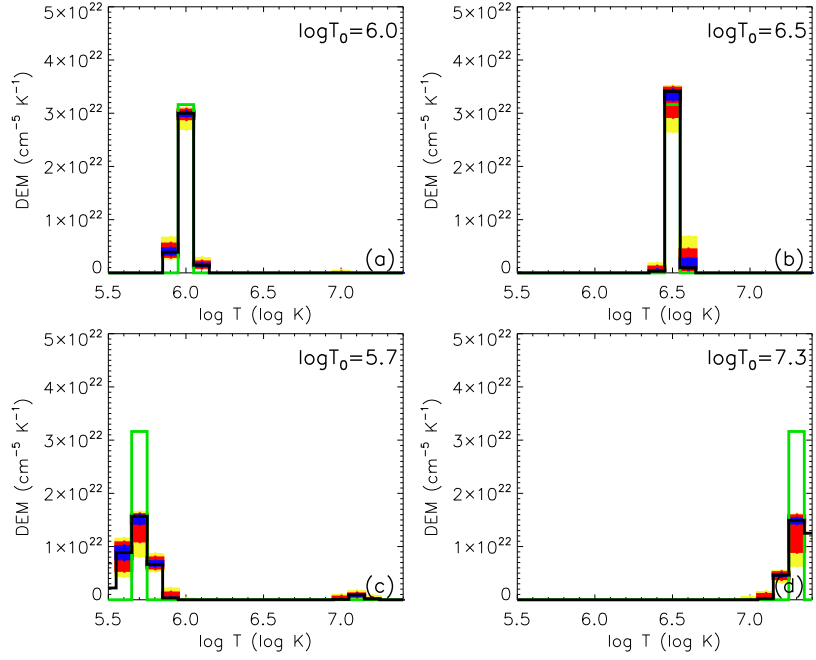}
               }
\vspace{0.0\textwidth}   
\caption{Reconstruction of single isothermal DEMs with $\log T$=6.0 (a), 6.5 (b), 5.7 (c), and 7.3 (d). The green solid lines show the model DEMs, the black solid lines display the best-fitted DEMs, and the colored rectangles have the same meaning as in Figure \ref{f3}.} \label{f14}
\end{figure}

In the single isothermal cases (Figure \ref{f14}), we find that in the temperature range of $\log T_{0}$=5.8--7.2, the input temperature is successfully located, with only minor ``spreading" into adjacent temperature bins. Since the input delta function DEM is fitted with splines, some EM ``spillage" is to be expected, and the results are still quite good. The full width at half maximum (FWHM) of the recovered DEM in this $T_0$ range is equal to 0.2 everywhere, but increases to 0.4 for $\log T_{0}$ out of this range.

\begin{figure} 
     \vspace{-0.0\textwidth}    
     \centerline{\hspace*{0.00\textwidth}
               \includegraphics[width=0.7\textwidth,clip=]{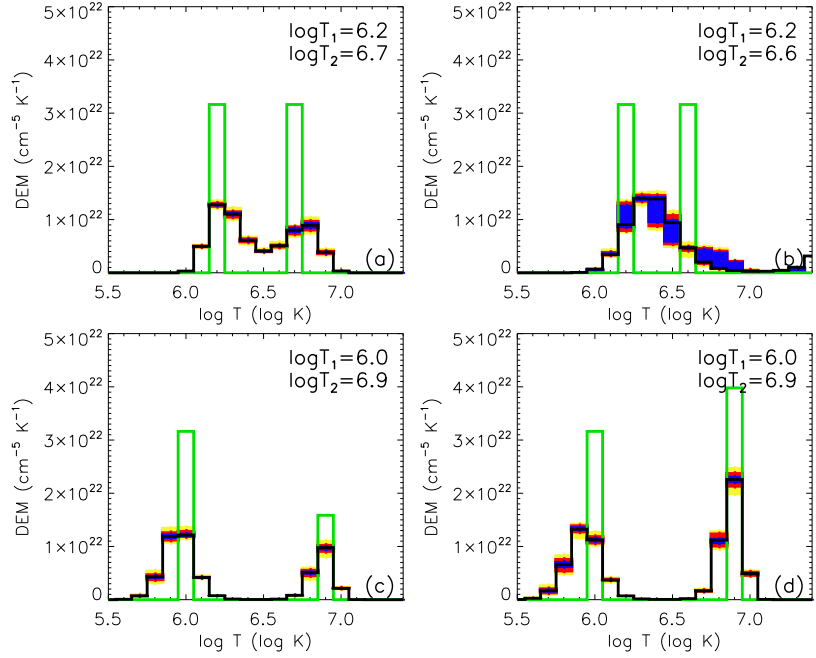}
               }
\vspace{0.0\textwidth}   
\caption{Reconstruction of double isothermal DEMs with with varying temperature separations and the DEM peaks. The green solid lines, the black solid lines, and the colored rectangles have the same meaning as in Figure \ref{f14}.} \label{f15}
\end{figure}

In the double isothermal cases (Figure \ref{f15}), we find that the fit EMs again always show some spread around the input temperatures $\log T_{1}$ and $\log T_{2}$, and the FWHM tends to be larger ($\geq 0.2$). The two DEM peaks can not be resolved when the peak separation $\Delta \log T$ ($\log T_{1} - \log T_{2}$) $\leq$ 0.4 (Figure \ref{f15}(b)). Nevertheless, the code still finds the DEM peak positions well once $\Delta \log T > 0.4$, independent of the $EM_{1}/EM_{2}$ ratio (Figure \ref{f15}(a), (c), and (d)). The output parameters of fitting a double Gaussian profile to the double isothermal DEM for four specific cases are listed in Table \ref{tb2}. The recovered DEM curves can be always fitted successfully by the double Gaussian profiles with fixed $\log T_{1}$ and $\log T_{2}$ as same as modeled $\log T_{1}$ and $\log T_{2}$ except the case b in Figure \ref{f15}.

\begin{figure} 
     \vspace{-0.0\textwidth}    
     \centerline{\hspace*{0.00\textwidth}
               \includegraphics[width=0.7\textwidth,clip=]{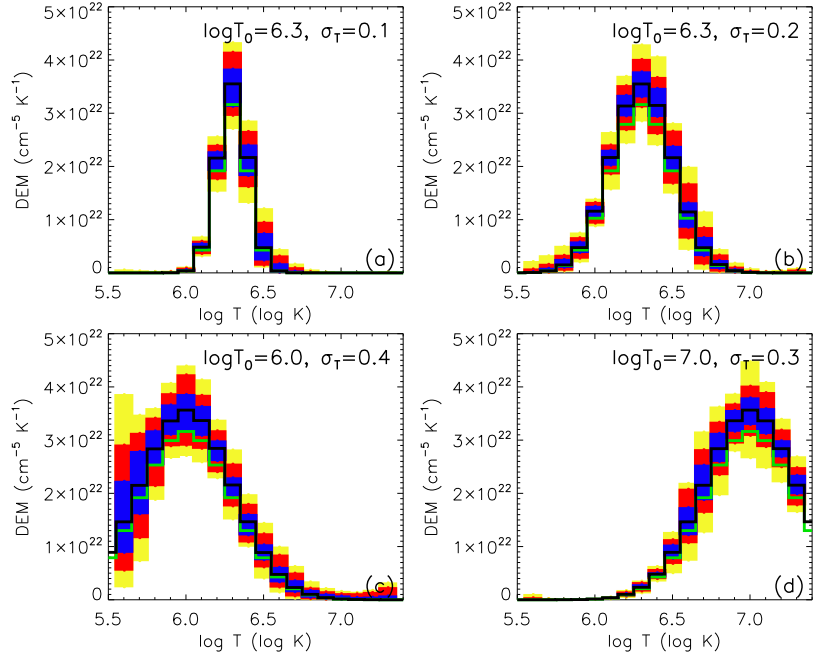}
               }
\vspace{0.0\textwidth}   
\caption{Reconstruction of single Gaussian DEMs with different $\log T_{0}$ and $\sigma_{T}$. The green solid lines, the black solid lines, and the colored rectangles have the same meaning as in Figure \ref{f14}.} \label{f16}
\end{figure}

\begin{figure} 
     \vspace{-0.0\textwidth}    
     \centerline{\hspace*{0.00\textwidth}
               \includegraphics[width=0.7\textwidth,clip=]{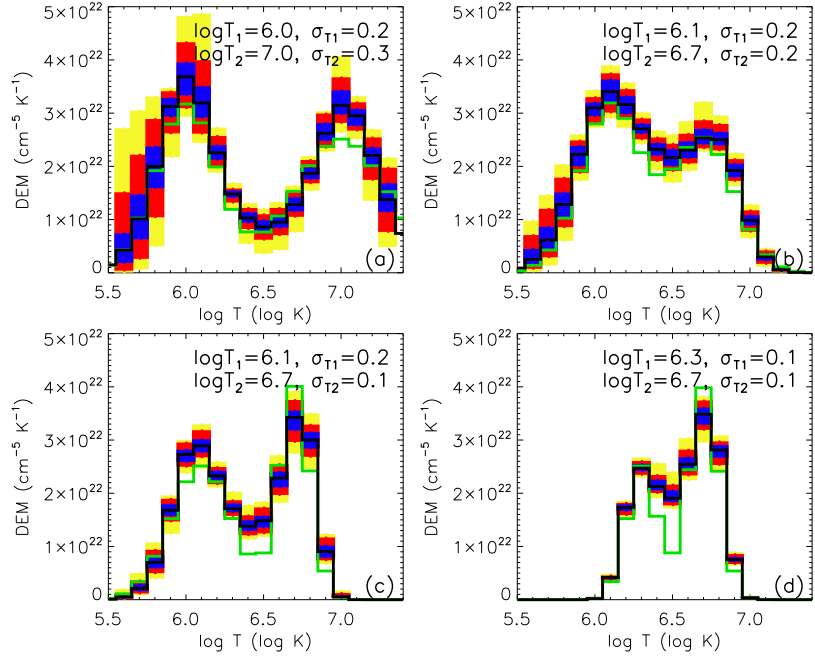}
               }
\vspace{0.0\textwidth}   
\caption{Reconstruction of double Gaussian DEMs with different temperature separations, DEM peaks, and $\sigma_{T}$. The green solid lines, the black solid lines, and the colored rectangles have the same meaning as in Figure \ref{f14}.} \label{f17}
\end{figure}

In the single Gaussian DEM cases (Figure \ref{f16}), we find that for any $\log T_{0}$ in the range of 5.7--7.3 with an input width $\sigma_{T} \geq$ 0.1, the best fitted DEMs are very close to the model Gaussian DEMs. The normalized error $\Delta = \Sigma (|DEM_i - DEM^{model}_{i} |)/\Sigma DEM^{model}_{i}   \leq 0.12$, the linear correlation coefficient $CC \geq 0.99$, and the chi-squared statistic $\chi^{2} = \Sigma (DEM_i - DEM^{model}_{i} )^2/\sigma^2 (DEM_{i} ) \leq 1.60$ (see Table \ref{tb3} for metrics for the cases in Figure \ref{f16}). We also note that the 100 MC solutions show little scatter around the best-fit solution, indicating the emission measures are fit with low uncertainty. Similar to the single Gaussian cases, with $\log T$ in the range of 5.7--7.3, any $\Delta \log T$ separation, and $\sigma_{T}$, the best fitted DEMs recover the double Gaussian DEMs profiles successfully (Figure \ref{f17}). The metrics of the goodness-of-fit for model DEMs are shown in Table \ref{tb3}. However, it is noted that the fit quality deteriorates when $\sigma_{T_{1}}$ and/or $\sigma_{T_{2}}$ = 0.1 (Figure \ref{f17}(c) and (d)). The normalized error $\Delta \geq 0.19$, the linear correlation coefficient $CC \leq 0.97$, the $\chi^{2}$ increases to be 40.75 for the case d of Figure \ref{f17}, for example.

\begin{figure} 
     \vspace{-0.0\textwidth}    
     \centerline{\hspace*{0.0\textwidth}
               \includegraphics[width=0.7\textwidth,clip=]{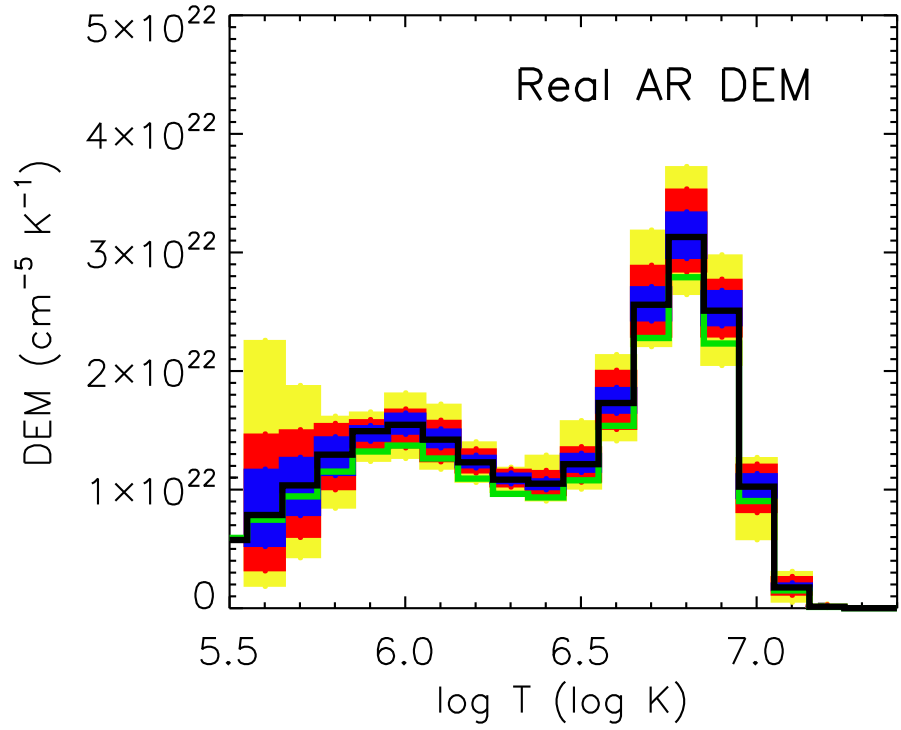}
               }
\vspace{0.0\textwidth}   
\caption{Reconstruction of the active region DEM from CHIANTI file ``active\b{ }region\b{ }oso6.dem". The green solid lines, the black solid lines, and the colored rectangles have the same meaning as in Figure \ref{f14}.} \label{f18}
\end{figure}

Finally, the CHIANTI active region DEM ``active\b{ }region\b{ }oso6.dem" is accurately reproduced across the entire temperature range with quite good accuracy (Figure \ref{f18}). The normalized error $\Delta = 0.12$, the linear correlation coefficient $CC = 0.99$, and the chi-squared statistic $\chi^{2} = 4.37$. Therefore, these tested cases suggest that ``xrt\b{ }dem\b{ }iterative2.pro" is a reliable routine that can reconstruct the DEMs with AIA data well, with the caveats that it has more difficulty with isothermal plasmas, and with separating nearby EM peaks. This is likely due to its use of smooth splines to define the DEM, which naturally have difficulty with extremely sharp features. It is worth mentioning that \citet{schmelz09a,schmelz09b} compared `xrt\b{ }dem\b{ }iterative2.pro" with the Markov-Chain Monte Carlo (MCMC) based DEM reconstruction algorithm by \citet{kashyap98}, and found that the two methods are in good agreement. The MCMC method, since it searches $\chi^2$ space more thoroughly, generally finds slightly lower $\chi^2$ minima, and ``spikier" solutions, but at the cost of considerably more computing time.

\bibliographystyle{apj}

\clearpage

\clearpage

\begin{table}
\caption{Properties of CME flux ropes, LFs, and dimming regions.}
\label{tb}
\begin{tabular}{ccccccc}
\\ \tableline \tableline
 Event         &  Region     & $\bar{T}$       & $\delta\log T^a$  & Maximum DEM     & Total EM   & Density       \\
               &             & (MK)            &        & ($\times$ 10$^{20}$ cm$^{-5}$ k$^{-1}$) &($\times$ 10$^{27}$ cm$^{-5}$) &($\times$ 10$^{9}$ cm$^{-3}$)   \\
\hline
\multicolumn{7}{c}{Flux Rope}\\
\hline
               & a           &8.3              &0.9      &3.6        &2.08       &  0.72             \\
CME1           & b           &9.1              &1.3      &8.8        &9.05       &  1.22             \\
               & c           &9.0              &1.3      &10.2       &10.40      &  1.32             \\

               & a           &10.9             &0.4      &2.6        &1.31       &  0.74             \\
CME2           & b           &9.5              &0.9      &7.4        &4.74       &  1.18             \\
               & c           &9.0              &1.3      &15.8       &13.70      &  2.38             \\

               & a           &6.8              &0.9      &24.5       &12.00     &  1.72             \\
CME3           & b           &9.0              &1.3      &1.2        &1.41      &  0.53             \\
               & c           &4.5              &0.4      &3.9        &2.43      &  0.59             \\
\hline
\multicolumn{7}{c}{Leading Front}\\
\hline
CME1           &  --         &2.1              &0.9      &7.9        &0.94      &  0.24             \\

CME2           &  --         &1.7              &0.9      &2.8        &0.26      &  0.11             \\

CME3           &  --         &1.9              &0.9      &0.5        &0.06      &  0.05             \\
\hline
\multicolumn{7}{c}{Dimming Region}\\
\hline
CME1           &  --         &2.2              &1.3      &3.9        &0.67      &  0.23             \\

CME2           &  --         &1.9              &0.9      &3.8        &0.44      &  0.18             \\

CME3           &  --         &1.8              &1.3      &0.7        &0.08      &  0.07             \\
\tableline
\end{tabular}

\vspace{0.03\textwidth}
$^a$ Width of the DEM curve at 10\% of the peak value based on a single Gaussian fit to the DEM curve.\\

\end{table}

\begin{table}
\caption{Parameters of fitting a double Gaussian profile to recovered double isothermal DEM.}
\label{tb2}
\begin{tabular}{ccccccc}
\\ \tableline \tableline
 Case        &  $\log T_{1}$ & $\sigma_{T_{1}}$  & $\log T_{2}$  &$\sigma_{T_{2}}$    &$R_{1}^a$   & $R_{2}^b$ \\
 \hline
 a            & 6.2           &0.2                  &6.7            &0.2                &  1.3       &  0.5    \\
 b            & 6.3           &0.3                  &6.5            &0.3                &  1.4       &  0.5    \\
 c            & 6.0           &0.2                  &6.9            &0.2                &  1.1       &  0.9    \\
 d            & 6.0           &0.4                  &6.9            &0.2                &  1.1       &  0.8    \\
\tableline

\end{tabular}

\vspace{0.03\textwidth}
$^a$ Ratio of the EM integrated with the recovered DEM to with the input DEM for the first isothermal peak.\\
$^b$ Ratio of the EM integrated with the recovered DEM to with the input DEM for the second isothermal peak.\\
\end{table}

\begin{table}
\caption{Metrics for goodness of fitted DEMs (Figures \ref{f16}--\ref{f18}).}
\label{tb3}
\begin{tabular}{cccccc}
\\ \tableline \tableline
 Case        &  $\Delta^a$ & $CC^b$  & $\chi^{2}$$^c$ & $R_{1}$  & $R_{2}$ \\
\hline
\multicolumn{6}{c}{Single Gaussian}\\
\hline
 a            & 0.12           &0.99                  &0.42     & 1.12   &    --       \\
 b            & 0.12           &0.99                  &1.14     & 1.12   &    --       \\
 c            & 0.12           &0.99                  &1.07     & 1.12   &    --       \\
 d            & 0.12           &0.99                  &1.60     & 1.12   &    --       \\
\hline
\multicolumn{6}{c}{Double Gaussian}\\
\hline
 a            & 0.15           &0.98                  &5.72     & 1.15  &   1.05       \\
 b            & 0.14           &0.99                  &8.20      & 1.15  &   1.07       \\
 c            & 0.20           &0.97                  &12.01      & 1.14  &   0.99       \\
 d            & 0.19           &0.97                  &40.75     & 1.08  &   0.99       \\
\hline
\multicolumn{6}{c}{Real DEM}\\
\hline
--            & 0.12           &0.99                  &4.37       & 1.12  & --           \\
\tableline
\end{tabular}

\vspace{0.03\textwidth}
$^a$ Normalized error $\Delta = \frac {\sum | DEM_{i}-DEM_{i}^{model} | }{ \sum DEM_{i}^{model}}$. \\
$^b$ Linear Pearson correlation coefficient. \\
$^c$ Chi-squared statistic $\chi^{2} = \sum \frac {(DEM_{i}-DEM_{i}^{model})^2}{\sigma^{2}(DEM_{i})}$.\\
\end{table}

\clearpage
\end{document}